\documentclass[onecolumn]{aa}
\usepackage{graphicx}
\usepackage{txfonts}
\usepackage{natbib}
\bibpunct{(}{)}{;}{a}{}{,}

\begin{document}

\title{Significance in gamma-ray astronomy -\\ the Li \& Ma problem in Bayesian
statistics}

\author{
S. Gillessen \and H.L. Harney}
\institute{MPI f\"ur Kernphysik, P.O.Box 103980, 69029 Heidelberg, Germany}
\offprints{Stefan.Gillessen, \email Stefan.Gillessen@mpi-hd.mpg.de}

\date{Received December 12, 2003; accepted September 15, 2004}

\abstract{
The significance of having detected an astrophysical
gamma ray source is usually calculated by means of a formula derived
by Li \& Ma in 1983. We solve the same problem in
terms of Bayesian statistics, which provides a logically more satisfactory
framework. We do not use any subjective 
elements in the present version of Bayesian statistics.
We show that for large count numbers and a weak source the Li \& Ma formula
agrees with the Bayesian result. For other cases the two
results differ, both due to the mathematically different
treatment and the fact that only Bayesian inference 
can take into account prior knowldege.
}

\authorrunning{S. Gillessen \& H.L. Harney}
\titlerunning{The Li \& Ma problem in Bayesian statistics}

\maketitle

\section{Introduction}

Consider an astronomical gamma ray observation 
aiming to detect a source. 
The existence of a source in a so-called on-region is
judged by the count number $N_{\mathrm{on}}$ originating from that region.
The counts in it are due to a possible source and the 
background. The latter
is determined by the count number $N_{\mathrm{off}}$ 
in some off-region. It must be chosen in such a way that 
one can exclude a priori that it 
contains a source. Hence, we use a
physically motivated choice of on- and off-regions
and not a blind search.
One also knows
the expected ratio $\alpha$ of the count numbers if
there is no source in the on-region.
The number $\alpha$ is given by the ratio of the sizes of 
the two regions, the ratio of the exposure times for both regions and the respective
acceptances:
\begin{equation}
\alpha = \frac{\kappa_{\mathrm{on}} \cdot t_{\mathrm{on}} \cdot A_{\mathrm{on}}}
{\kappa_{\mathrm{off}} \cdot t_{\mathrm{off}} \cdot A_{\mathrm{off}}} \,.
\end{equation}
Given ($\alpha$, $N_{\mathrm{on}}$, $N_{\mathrm{off}}$) the question is how significantly a 
possible source has been detected. 
A positive identification obviously requires 
$ N_{\mathrm{on}} > \alpha \, N_{\mathrm{off}}$. \citet{lim83} 
discuss several
possible estimates of the significance.
Estimating it as the ratio of excess counts above background to 
the background's standard deviation yields (\citealp[eq.~(5)]{lim83})
\begin{equation}
S_{\mathrm{LM}\,1} = \frac{N_{\mathrm{on}} - \alpha N_{\mathrm{off}}}{\sqrt{N_{\mathrm{on}} + \alpha^2 N_{\mathrm{off}}}}\,.
\label{slm1}
\end{equation}
However, one could as well argue that the desired measure of 
significance should correspond to the probability
that all counts were due to the background. That yields 
(\citealp[eq.~(9)]{lim83}):
\begin{equation}
S_{\mathrm{LM}\,2} = \frac{N_{\mathrm{on}} - \alpha N_{\mathrm{off}}}{\sqrt{\alpha (N_{\mathrm{on}} + N_{\mathrm{off}})}} \,.
\label{slm2}
\end{equation}
Li \& Ma argue that for $\alpha < 1$, $S_{\mathrm{LM}\,1}$ underestimates 
the significance, $S_{\mathrm{LM}\,2}$ overestimates it. They finally 
advocate the significance $S_{\mathrm{LM}}$ (\citealp[eq.~(17)]{lim83}) in the form
\begin{equation}
S_{\mathrm{LM}} = \sqrt{2} \left(N_{\mathrm{on}} \cdot \ln \left( 
\frac{(1+\alpha)\,N_{\mathrm{on}}}{\alpha \, (N_{\mathrm{on}} + N_{\mathrm{off}})} \right) 
+ N_{\mathrm{off}} \cdot \ln \left( \frac{(1+\alpha) N_{\mathrm{off}}}{N_{\mathrm{on}} + N_{\mathrm{off}}} \right)
\right) ^{1/2} \,.
\label{lima}
\end{equation}
As a function of the random variables $N_{\mathrm{on}}$ and $N_{\mathrm{off}}$ this is
itself a random variable.
If no source is present this variable is nearly normally distributed
even for small count numbers (according to the authors for $N_{\mathrm{on}} , \, N_{\mathrm{off}} \gtrsim 10$).
For a single measurement (given by the numbers $\alpha$, $N_{\mathrm{on}}$ and $N_{\mathrm{off}}$)
one can interpret $S_{\mathrm{LM}}$ as statistical significance.
The argument of Li \& Ma hinges on the fact that $S_{\mathrm{LM}}$ has
a normal distribution. They have tested this by Monte Carlo methods.

In the present paper we define and evaluate the significance $S_{\mathrm{B}}$
of the existence of a source in terms of Bayesian statistics. We do so for
several reasons.
\begin{itemize}
\item We consider Bayesian statistics to provide a logically more satisfactory
inference than the arguments of classical statistics used by Li \& Ma.
\item Bayesian significance does not leave a choice between
several definitions of significance. We do not consider the prior
distribution to be a subjective element in statistical inference,
nor do we take it to be uniform either. Rather we define it by a formal
rule which is based on a symmetry principle. This may be called an
objective Bayesian approach.
\item Bayesian statistics do not require a random variable that has
an approximately normal distribution. Bayesian inference is therefore 
valid for any count number. It does not require verification by
Monte Carlo methods.
\end{itemize}

The classical significance $S_{\mathrm{LM}}$ and the Bayesian significance $S_{\mathrm{B}}$
do not have the same meaning. 
The first expresses a probability that the
assumption ''there is no source'' conflicts with observation. 
The corresponding test function can be defined in various ways. The second
expresses the probability that the intensity of the source is larger
than zero. This probability is taken from a posterior distribution 
of the intensity parameter, which is a well-defined result of 
Bayesian inference.
Although the two quantities do not have the same meaning, we compare
the numerical values because the application of Bayesian statistics is
not common practice and there
is a limiting situation in which both values agree. It occurs in
the frequent case when the source is weak and the count numbers
are high.

\section{Basics of Bayesian statistics}
\subsection{Problems depending on one parameter}
\label{basics}
Bayesian statistics provides a way to infer
physical parameters from observed data. The dependence
of the observed quantities on the parameters is
statistical. Hence, it is described in terms of
probability distributions.
In the following we shall use the Poisson distribution
\begin{equation}
p_{\mathrm{P}}(n|\lambda) =  \frac{\lambda^n}{n!} \, e^{-\lambda} 
\label{poisson}
\end{equation}
and the binomial distribution
\begin{equation}
p_{\mathrm{B}}(n|\lambda;N) = {N \choose n} \lambda^n (1-\lambda)^{N-n}\,.
\label{binomial}
\end{equation}
The parameter is a real number $\lambda$, 
the observed datum is a whole number $n$.
In order to derive the parameter, 
the conditional distribution $p(n|\lambda)$
must be proper so that
\begin{equation}
\sum_n p(n|\lambda) = 1 \,.
\label{proper}
\end{equation}
The Poisson and the binomial models are proper. 
The probability for the parameter to have the value $\lambda$ 
is found by means of Bayes' theorem\footnote{For improper models the prior
distribution needed in Bayes' theorem is not defined.}:
\begin{equation}
P(\lambda | n) = \frac{p(n | \lambda) \, \mu(\lambda)}
{\int p(n | \lambda') \, \mu(\lambda')\, d\lambda'} \,.
\label{bayes}
\end{equation}
The posterior distribution $P(\lambda | n)$ contains the information one can
deduce from the data. It is a distribution of the parameter given the data
whereas the model $p(n|\lambda)$ is a distribution of the data given
the parameter.\\
Bayes' theorem does not determine 
the so-called prior distribution $\mu(\lambda)$ in equation (\ref{bayes}). However,
demanding in addition a symmetry for the model yields the
prior distribution: In order to ensure an unbiased inference
of $\lambda$ in the sense that the information obtained on 
$\lambda$ does not depend on the actually true value of $\lambda$,
one demands that the distribution is form-invariant. This means that
there is a group of transformations that relates the observable $n$ 
to the parameter 
$\lambda$. The measure of the group can then be identified with
the prior distribution in equation (\ref{bayes}), see Harney (2003), chap.~6.
The measure of the group is obtained 
by ''Jeffreys' rule'' \citep[see][chap.~3]{jeffreys}:
\begin{eqnarray}
\mu(\lambda) & = & 
\left< \left(\partial_{\lambda} \ln p(n | \lambda) \right)^2 
\right>_p^{1/2} \, .
\label{jeffrey}
\end{eqnarray}
Here, $\left< f(n) \right>_p$ denotes the expectation value of $f$ with respect to the
distribution $p$.
For the evaluation of the right hand side of 
equation (\ref{jeffrey}),  
see sect. \ref{calculationofmeasures}. Under a transformation of 
the parameters, the measure transforms with the Jacobian
of the transformation, so that any derived probabilities are
not affected by a reparameterization. The measure $\mu$ is
not necessarily a proper distribution. One must only demand
that the normalizing integral in equation~(\ref{bayes})
exists and thus the posterior distribution is proper.

One is usually interested in an error interval for the derived
value of the parameter $\lambda$.
It can be constructed as a Bayesian interval: Given a
preselected probability $K$,
it is the shortest interval $[\lambda_1,\lambda_2]$ for which
\begin{equation}
\int_{\lambda_1}^{\lambda_2} P(\lambda | n) \, d\lambda = K \,.
\end{equation}
It can be shown (\citealp[see][chap.~3]{harney})
that if the Bayesian interval is unique, it is defined by 
some constant $C(K)$ such that the
interval contains the points for which
\begin{equation}
\frac{P(\lambda|n)}{\mu(\lambda)} > C(K) \,.
\label{hoehenlinie}
\end{equation}
With (\ref{bayes}) one sees that
$C(K)$ is the level of a contour line of the model
$p(n|\lambda)$ taken as function of $\lambda$.\\
For the problem at hand we need the probability 
that the Bayesian interval excludes some lower
bound $\lambda_{\mathrm{min}}$. This can be calculated from the posterior distribution
in two steps: 
\begin{itemize}
\item Find the corresponding Bayesian interval. The lower bound is
$\lambda_{\mathrm{min}}$, the upper bound $\lambda_{\mathrm{up}} > \lambda_{\mathrm{min}}$ is
found by solving the equation:
\begin{equation}
p(n|\lambda_{\mathrm{up}})= p(n|\lambda_{\mathrm{min}})\,.
\label{eqn}
\end{equation}
\item The probability is then 
\begin{equation}
K(\lambda > \lambda_{\mathrm{min}}) = \int_{\lambda_{\mathrm{min}}}^{\,\lambda_{\mathrm{up}}} 
P(\lambda' | n) \, d\lambda' \,
\label{KK}
\end{equation}
as any $K$ bigger than that would yield a Bayesian interval
that includes $\lambda_{\mathrm{min}}$.
\end{itemize}
For $K$ close to unity it is handy to express it in a different,
highly non-linear scale,
which we call significance $S$. The conversion is done by
\begin{equation}
\mathrm{erf}\left(\frac{S}{\sqrt{2}}\right) = K(\lambda > \lambda_{\mathrm{min}}) \,, 
\label{k2s}
\end{equation}
where the error function is defined by
\begin{equation}
\mathrm{erf}\left(\frac{S}{\sqrt{2}}\right) = 
\frac{1}{\sqrt{2 \pi}} \, \int_{-S}^{\,S} e^{-x^2/2} \, dx \,.
\label{erfdef}
\end{equation}
This yields the significance in the Bayesian context.
Note that the term significance is used here in a sense that can be read as
'if the posterior distribution were Gaussian, 
the probability would correspond to $S$ standard deviations'. A short-hand
form of that is 'the significance is S sigma'.
It is not required that the posterior distribution is Gaussian. However,
the definition~(\ref{k2s}) is motivated by the fact that for large
count numbers the posterior distribution does approach
a Gaussian.\\
The error function in equation (\ref{erfdef}) is odd. For sufficently large $S$ 
it can be approximated by
\begin{equation}
\mathrm{erf}\left(\frac{S}{\sqrt{2}}\right) \approx
1 - \sqrt{\frac{2}{\pi}} \cdot 
\frac{1}{S} \cdot \exp\left(-\frac{S^2}{2}\right) \, ,
\,\,\,\,(S^2 \gg 1, \, S > 0) \, .
\label{erfapprox}
\end{equation}
\subsection{Reducing multi-parametric problems}
\label{marginalization}
The appropriate model may depend on more parameters than are
interesting. That means that one has to integrate over the uninteresting
parameters. 
The question arises whether one should integrate
first and apply Bayes' theorem then or if the integration
should be performed after the application of Bayes' theorem.
The second way (obtaining the full
posterior distribution first and integrating afterwards) 
does not provide the measure of the interesting
parameters only, although this measure 
is needed to find the Bayesian interval
via equation (\ref{hoehenlinie}).
This difficulty is related to the marginalization paradox
\footnote{Even if
the full measure factorizes into two factors, one depending
only on the interesting parameters and the other only on the uninteresting 
ones, the factors need not be meaningful
measures for the minor-dimensional problem \citep{bernardo}. An example
can be found in \citealp[chap.~12.1]{harney}}
\citep{dawid}. \\
Thus it is reasonable to go to a minor model
before applying Bayes' theorem. If the final minor model
has only one parameter, one can apply
the methods from sect. \ref{basics}.\\
The minor model which one constructs by integration shall be
invariant under a transformation of the integrated parameters. Thus
one needs the conditional measure in the integration kernel. It
is obtained by Jeffreys' rule if one
considers the interesting parameters as fixed. The minor model
$q(n|\lambda_1)$ for a model $p(n|\lambda_1,\,\lambda_2)$ is thus given by
\begin{equation}
q(n|\lambda_1) = \int p(n|\lambda_1,\,\lambda_2) \mu(\lambda_2|\lambda_1) 
\, d\lambda_2\,.
\end{equation}

\section{Solution by means of Bayesian statistics}
\label{bayessol}
The expected count number $\lambda_{\mathrm{on}}$ in the on-region is 
due to both background counts and 
the possible existence of a source. With the expected count number $\lambda_{\mathrm{off}}$ in
the off-region and the expected count number $\lambda_\mathrm{s}$ from
the source, one has
\begin{equation}
\lambda_{\mathrm{on}} = \alpha \, \lambda_{\mathrm{off}} + \lambda_{\mathrm{s}} \,,
\label{l0}
\end{equation}
since the expectation values linearly depend upon the intensities.
\subsection{The problem in its original parameters}
The probability of observing $N_{\mathrm{on}}$ and $N_{\mathrm{off}}$
given the independent parameters $\lambda_{\mathrm{on}}$ and 
$\lambda_{\mathrm{off}}$
is the product of the Poisson distributions:
\begin{equation}
p_0(N_{\mathrm{on}}, N_{\mathrm{off}}| \lambda_{\mathrm{on}}, \lambda_{\mathrm{off}}) 
=p_P(N_{\mathrm{on}}|\lambda_{\mathrm{on}}) \cdot  
p_P(N_{\mathrm{off}}|\lambda_{\mathrm{off}}) \,.
\label{p0}
\end{equation}
From this distribution one wants to infer the confidence level
to which $\lambda_{\mathrm{s}} = 0$ can be excluded. 
Hence, $\lambda_{\mathrm{s}}$ must be one of the parameters of the model.
Going to the parameters $(\lambda_{\mathrm{s}},\lambda_{\mathrm{off}})$ 
does not change any of the measures, as the transformation
(eq. (\ref{l0})) has the Jacobian $1$. One only has to read $\lambda_{\mathrm{on}}$
as $\lambda_{\mathrm{on}}(\lambda_{\mathrm{s}}, \, \lambda_{\mathrm{off}})$.
The parameter $\lambda_{\mathrm{off}}$ is not
interesting, and one has to integrate over it as discussed
in sect. \ref{marginalization}.
Thus the natural choice seems to be
\begin{equation}
q_0(N_{\mathrm{on}}, N_{\mathrm{off}}| \lambda_{\mathrm{s}}) = \int p_0(N_{\mathrm{on}}, N_{\mathrm{off}}| \lambda_{\mathrm{s}}, \lambda_{\mathrm{off}}) \, 
\mu_0(\lambda_{\mathrm{off}}|\lambda_{\mathrm{s}}) \, d\lambda_{\mathrm{off}} \, .
\label{q}
\end{equation}
The conditional measure $\mu_0(\lambda_{\mathrm{off}}|\lambda_{\mathrm{s}})$ 
is calculated in eq.~(\ref{measuremu0}).
Unfortunately $q_0$ is an improper model since $\mu_0$
is not integrable (see sect. \ref{evalq}).
This problem is somewhat unexpected. It is a consequence of the
fact that the measure of the Poisson model (see section~\ref{measurepoisson})
is improper.
\subsection{Transformation to a proper model}
However, a simple transformation circumvents the problem. We define
\begin{eqnarray}
\Lambda & = & \lambda_{\mathrm{on}} + \lambda_{\mathrm{off}} \,, \nonumber \\
\omega & = & \frac{\lambda_{\mathrm{on}}}{\Lambda} \,,\nonumber \\
N & = & N_{\mathrm{on}} + N_{\mathrm{off}}\,.
\label{trafo}
\end{eqnarray}
The parameter $\omega$ represents the fraction of the total
intensity $\Lambda$ in the on-region and has the boundaries
\begin{equation}
\omega_{\mathrm{min}}= \frac{\alpha}{1+\alpha} \leq \omega \leq 1 \,.
\label{omegabounds}
\end{equation}
Since one is free to choose the units in which the intensities are measured, 
the problem can only depend on the relative intensities. 
This freedom of gauge becomes transparent in the new parameters. 
The significance can only depend on $\omega$,
the total count number $N$ only on the uninteresting parameter
$\Lambda$.
When one introduces the new parameters $\omega$ and $\Lambda$ 
into equation (\ref{p0}) one sees explicitly that they are independent, since
the model $p_0$ factorizes in the new parameters (see eq. (\ref{p0t}))
according to
\begin{equation}
p_0(N_{\mathrm{on}}, N_{\mathrm{off}}| \lambda_{\mathrm{s}}, \lambda_{\mathrm{off}}) 
= p_{\mathrm{P}}(N|\Lambda) \cdot p_{\mathrm{B}}(N_{\mathrm{on}}|\omega;N).
\end{equation}
The total count number is given by Poisson statistics, the
subdivison of the counts into on- and off-regions, given
a certain $\omega$, is governed by the binomial distribution.
Therefore we infer $\omega$ from the binomial model 
only and consider the total count number $N$ as fixed. In other words,
we do not normalize $p_{\mathrm{B}}(N_{\mathrm{on}}|\omega;N)$
with respect to $N$. Then $p_{\mathrm{B}}$ is proper.
The measure $\mu_{\mathrm{B}}(\omega)$ of $p_{\mathrm{B}}$ is proper (see eq. (\ref{measureofbinomial})):
\begin{eqnarray}
\mu_{\mathrm{B}}(\omega)&=& \left( \frac{N}{\omega (1-\omega)} \right)^{1/2} \, .
\label{mu1}
\end{eqnarray}

\subsection{Explicit solution}
One can safely apply Bayes' theorem to $p_{\mathrm{B}}$ to obtain
\begin{equation}
P_1(\omega|N_{\mathrm{on}};N) =\frac
{p_{\mathrm{B}}(N_{\mathrm{on}}|\omega;N) \cdot \mu_{\mathrm{B}}(\omega)}
{{\cal N}_1} \,.
\end{equation}
The normalization ${\cal N}_1$ is 
\begin{eqnarray}
{\cal N}_1 &=& \int_{\omega_{\mathrm{min}}}^1 p_{\mathrm{B}}(N_{\mathrm{on}}|\omega;N) 
\cdot \mu_{\mathrm{B}}(\omega) \, d\omega \, \nonumber \\
& = & \sqrt{N} \, \cdot \frac{\Gamma\left(\frac{1}{2}+N_{\mathrm{on}}\right) \, 
\Gamma\left(\frac{1}{2}+N_{\mathrm{off}}\right) -
N! \cdot B_{\omega_{\mathrm{min}}}\left(\frac{1}{2}+N_{\mathrm{on}},\frac{1}{2}+N_{\mathrm{off}}\right) } 
{N_{\mathrm{on}}! \cdot N_{\mathrm{off}}!} \,,
\label{n1}
\end{eqnarray}
where $B_z(a,b)$ is the incomplete Beta function. 
Therewith the posterior distribution is:
\begin{equation}
P_1(\omega | N_{\mathrm{on}}; N) = 
\frac
{N! \cdot (1-\omega)^{(N_{\mathrm{off}}-1/2)}\cdot
\omega^{(N_{\mathrm{on}}-1/2)}}
{\Gamma\left(\frac{1}{2}+N_{\mathrm{on}}\right) \, \Gamma\left(\frac{1}{2}+N_{\mathrm{off}}\right) -
N! \cdot B_{\omega_{\mathrm{min}}}
\left(\frac{1}{2}+N_{\mathrm{on}},\frac{1}{2}+N_{\mathrm{off}}\right)}\,.
\label{aposteriori}
\end{equation}
\begin{figure}
\begin{center}
\includegraphics[width=12cm]{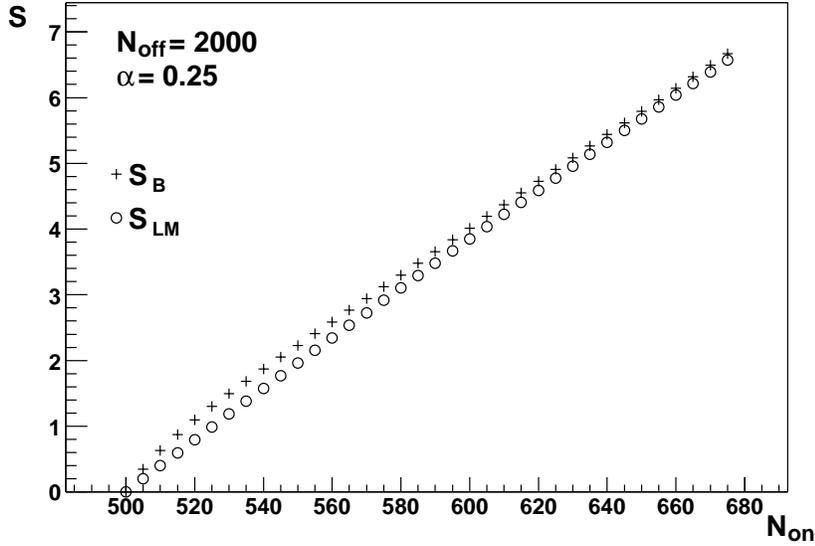}
  \caption{~Comparison of $S$ as a function of $N_{\mathrm{on}}$ for $\alpha=0.25$, $N_{\mathrm{off}}=2000$.
	  Significance $S_{\mathrm{LM}}$ according to
	 Li \& Ma (circles) and Bayesian significance $S_{\mathrm{B}}$ (crosses)
	 }
\label{fig1}  
\end{center} 
\end{figure}
\begin{figure}
\begin{center}
\includegraphics[width=12cm]{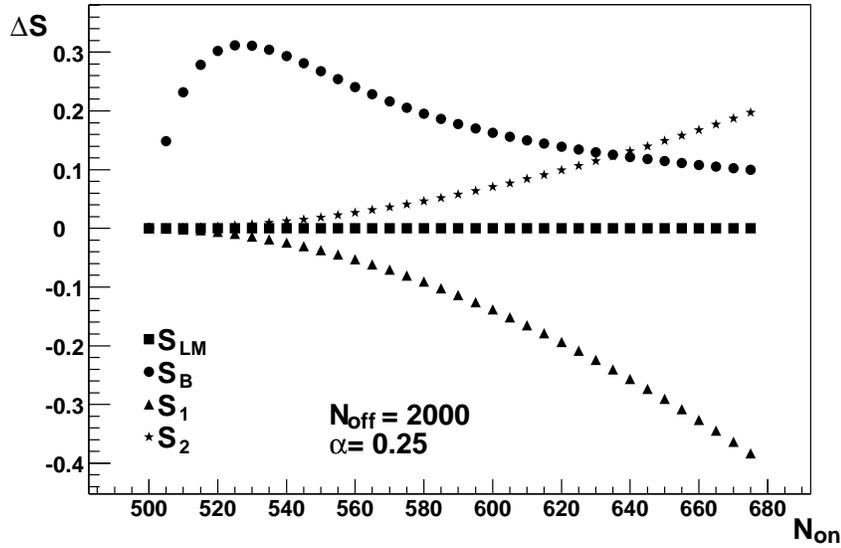}
  \caption{~Comparison of $S$ as function of $N_{\mathrm{on}}$ for $\alpha=0.25$, $N_{\mathrm{off}}=2000$.
         The difference $\Delta S = S - S_{\mathrm{LM}}$ is shown for
	 $S=S_{\mathrm{B}}$ (dots) and the two estimates $S=S_{LM\,1}$ (equation (protect{\ref{slm1}}), 
	 triangles) and $S=S_{LM\,2}$ 
	 (equation (\protect{\ref{slm2}}), stars).
	 }
\label{fig2}  
\end{center} 
\end{figure}
For the calculation of the significance one needs the integral 
over $P_1$:
\begin{eqnarray}
I_1(\omega) &= & \int_{\frac{\alpha}{\alpha+1}}^{\omega} P_1(v | N_{\mathrm{on}}; N)\, dv \nonumber \\
& = & \frac
{N! \cdot \left( B_{\omega}\left(\frac{1}{2}+N_{\mathrm{on}},\frac{1}{2}+N_{\mathrm{off}}\right) - 
B_{\omega_{\mathrm{min}}}\left(\frac{1}{2}+N_{\mathrm{on}},\frac{1}{2}+N_{\mathrm{off}}\right) \right)}
{\Gamma\left(\frac{1}{2}+N_{\mathrm{on}}\right) \, \Gamma\left(\frac{1}{2}+N_{\mathrm{off}}\right) 
- N! \cdot B_{\omega_{\mathrm{min}}}\left(\frac{1}{2}+N_{\mathrm{on}},\frac{1}{2}+N_{\mathrm{off}}\right)} 
\,.\nonumber \\
& &
\label{i1}
\end{eqnarray}
The probability that a source has been detected is given by
the probability that $\lambda_{\mathrm{s}} >0$. 
In the new parameters
one wants to determine the confidence level to which one can
exclude that $\omega$ equals its lower bound $\omega_{\mathrm{min}}$. Hence,
one must solve the equation
\begin{equation}
p_{\mathrm{B}}(\omega_{\mathrm{up}}) = p_{\mathrm{B}}(\omega_{\mathrm{min}}) \,.
\label{eqnbin}
\end{equation}
This cannot be solved analytically. 
However, one can prove
that for $N_{\mathrm{on}}, N_{\mathrm{off}} > 0$ exactly one solution $\omega_{\mathrm{up}} \ne \omega_{\mathrm{min}}$ exists,
since the binomial model has then a single maximum and no minima
(see sect. \ref{proof}).\footnote{If $N_{\mathrm{on}}=0$, any Bayesian interval includes $\omega_{\mathrm{min}}$
and one cannot - with any probability - affirm the existence of a source. The case $N_{\mathrm{off}}=0$ 
entails a Bayesian interval including $\omega=1$. Then 
one cannot affirm the absence of a source with any probability.} 
With $\omega_{\mathrm{up}}$ the significance is
\begin{equation}
S_{\mathrm{B}} = \sqrt{2} \cdot \mathrm{erf}^{-1}(I_1(\omega_{\mathrm{up}})) \, ,
\label{result}
\end{equation}
where $\mathrm{erf}^{-1}$ is the inverse of the error function.
Due to the appearance of $\omega_{\mathrm{up}}$ one cannot evaluate equation
(\ref{result}) any further. However, we can give a Mathematica script which
calculates the Bayesian significance $S_{\mathrm{B}}$ in the described way
(see sect. \ref{script}).
In figs. \ref{fig1} and \ref{fig2} the Bayesian significance is compared
to the Li \& Ma formula for a set of typical count
numbers.

\section{Large count numbers}
\subsection{Li \& Ma}
The procedure by Li \& Ma is designed for the case of large count 
numbers. This is explicitely mentioned in the their paper (\citealp{lim83}) and
it becomes apparent if one reparametrizes equation~(\ref{lima}) in the following two
variables:
\begin{eqnarray}
N_{\mathrm{BG}}&=&\alpha \, N_{\mathrm{off}} \\
r &=& \frac{N_{\mathrm{on}} - N_{\mathrm{BG}}}{N_{\mathrm{BG}}} \,\,.
\end{eqnarray}
Here, $N_{\mathrm{BG}}$ is the count number expected in the on-region when no
source is present and $r$ is the ratio of excess counts to the
expected background. A positive significance requires $r > 0$.
Expressing $S_{\mathrm{LM}}$ in the observables $(N_{\mathrm{BG}},r)$ gives
\begin{equation}
S_{\mathrm{LM}} = \sqrt{2 N_{\mathrm{BG}}} \, \left(
(1+r) \, \ln \frac{(1+r)(1+\alpha)}{1+\alpha+\alpha r} + \frac{1}{\alpha} 
\ln \frac{1+\alpha}{1+\alpha+\alpha r}
\right)^{1/2} \,\,.
\label{limar}
\end{equation}
Hence, $S_{\mathrm{LM}}$ grows proportional to $\sqrt{N_{\mathrm{BG}}}$ as one
would expect for significance. The point is that no other dependencies on
$N_{\mathrm{BG}}$ are present, as the rest of equation~(\ref{limar})
depends on the ratio of $r$ and $\alpha$ only. \\

\subsection{Bayes}
For the sake of comparison we must bring the Bayesian significance into the same
form, such that its dependence on $N_{\mathrm{BG}}$ is the same as for
$S_{\mathrm{LM}}$. That means that one has to take the limit of large $N_{\mathrm{BG}}$.\\
We can approximate 
the posterior distribution (eq. (\ref{aposteriori}))
by a Gaussian for large count numbers. The apparent advantage is that this
distribution can be treated analytically.
The approximation is done best 
in the parameter in which the measure is uniform.
Then the model and the
posterior distributions are proportional to each other.
Inspecting equation (\ref{mu1}) shows that this happens for
the parameter
\begin{equation}
\phi=\arcsin\left(\sqrt{\omega}\, \right)\,.
\end{equation}
The approximation is calculated in appendix~\ref{gaussapprox}.
Using $\phi_0 = \arctan(\sqrt{N_{\mathrm{on}}/N_{\mathrm{off}}})\,$ the result is
\begin{equation}
P_2(\phi|N_{\mathrm{on}};N)=\frac{1}{{\cal N}_2} 
\sqrt{\frac{2N}{\pi}} 
\exp\left(-2N \,(\phi-\phi_0)^2 \right) \,\,.
\label{P2}
\end{equation}
If ${\cal N}_2 = 1$, then $P_2$ is normalized
in $]-\infty,\infty[$.
The additional normalization factor ${\cal N}_2$ is due to the limited definition region
of $\omega$, which means that $\phi$ is limited to
\begin{equation}
\phi_{\mathrm{min}}=\arcsin\left(\sqrt{\omega_{\mathrm{min}}}\,\right) \leq 
\phi \leq \frac{\pi}{2} \,\,.
\end{equation}
It is handy to define the probability $K_2$ as if $\phi$ was defined on the entire real axis:
\begin{equation}
K_2 = \int_{\phi_{\mathrm{min}}}^{\phi_{\mathrm{up}}} 
P_2(\phi|N_{\mathrm{on}};N)\, d\phi \left| \begin{array}{l}\\
{\cal N}_2 = 1\end{array}\right. \,\,,
\end{equation}
The value of $K_2$ is
\begin{equation}
K_2=\mathrm{erf}\left(\sqrt{2 N} \,(\phi_0 - \phi_{\mathrm{min}}) \right) \,\,.
\end{equation}
The corresponding significance $S_2$ can easily be given as an analytical
expression:
\begin{eqnarray}
S_2 &=& 2 \sqrt{N} \, (\phi_0 - \phi_{\mathrm{min}}) \nonumber \\
&=& 2 \sqrt{N} \arcsin \frac{\sqrt{N_{\mathrm{on}}} - \sqrt{\alpha N_{\mathrm{off}}} }
{\sqrt{(1+\alpha) N}} \nonumber \\
&=& 2 \sqrt{N_{BG}} \sqrt{\frac{1+\alpha + \alpha r}{\alpha}} \,
\arcsin \frac{\sqrt{\alpha +\alpha r} - \sqrt{\alpha}}
{\sqrt{(1+\alpha)(1+\alpha + \alpha r)}} \,\,.
\label{s2}
\end{eqnarray}
The actual factor ${\cal N}_2$ will differ from unity. It is found by the condition
\begin{equation}
1=\int_{\phi_{\mathrm{min}}}^{\pi/2} P_2(\phi|N_{\mathrm{on}};N)\, d\phi \,\,.
\label{N2cond}
\end{equation}
For large count numbers 
the relevant range of $\phi$ is close to the position of the maximum, i.e. $\phi_0$. A
crucial property of $P_2$ is that it does not vanish at 
$\phi_{\mathrm{min}}$.
The value of $\phi_0$ is not far from $\phi_{\mathrm{min}}$. Therefore
one can show that the upper limit of the integration in 
equation~(\ref{N2cond}) can be replaced by infinity, as the corresponding
correction vanishes exponentially with growing count numbers.
Then one obtains
\begin{equation}
{\cal N}_2 = \frac{1}{2} \left(1 + K_2 \right) \,\,.
\label{n2}
\end{equation}
Note that ${\cal N}_2$ is close to unity, and it is necessarily smaller than
unity. 
With the additional normalization factor ${\cal N}_2$
the integration over $P_2$ gives the Bayesian probability $K_{\mathrm{B}}$ in
our approximation. Using the fact that $(1-K_2) \ll 1$ one gets
\begin{eqnarray}
K_{\mathrm{B}} &=& \frac{1}{{\cal N}_2} K_2 = \frac{2 K_2}{1+K_2} = \frac{1-(1-K_2)}{1-(1-K_2)/2}\nonumber\\
&\approx& \left(1-\frac{}{} (1-K_2)\right) \, \left(1 + \frac{1-K_2}{2}\right) \approx 1 - \frac{1-K_2}{2} = \frac{1+K_2}{2}
\end{eqnarray}
Going to the significance scale we have
\begin{equation}
\mathrm{erf} \left(\frac{S_{\mathrm{B}}}{\sqrt{2}}\right) = \frac{1}{2} \left( 
1+\mathrm{erf}\left(\frac{S_{2}}{\sqrt{2}}\right)
\right)
\end{equation}
Using equation (\ref{erfapprox}) one gets
\begin{equation}
\frac{1}{S_{\mathrm{B}}} \exp\left(-\frac{S^2_{\mathrm{B}}}{2} \right)
\approx \frac{1}{2}  \, \frac{1}{S_{2}} 
\exp\left(-\frac{S^2_{2}}{2} \right) \,.
\end{equation}
Setting $S_{\mathrm{B}}=(1+\delta) \, S_2$
and neglecting higher orders of $\delta$ yields
\begin{equation}
S_{\mathrm{B}}=S_{2} \cdot 
\left( 1 + \frac{\ln 2}{S_{2}^2} \right) \,\,.
\label{sbapprox}
\end{equation}
The second term in this formula is due to the limited definition region of the
source intensity parameter $\lambda$. With equation~(\ref{s2}) one sees that its 
contribution
becomes negligible for large $N_{BG}$ as it vanishes like $1/N_{BG}$. 
Then one simply has $S_{\mathrm{B}}=S_{2}$ which is plausible, as for large
count numbers the distribution will become more and more concentrated around
its maximum and therefore in the limit the definition region of the parameter
no longer has an effect. So $S_2$ is the Bayesian expression 
which can be compared to the Li \& Ma significance
as given in equation~(\ref{limar}). Apparently Bayesian inference and classical
statistics also then yield different estimates for the significance.

\section{Large count numbers and weak source}
Typically, in gamma ray astronomy the detected sources are at 
the limit of the instruments' sensitivities. Therefore 
long observation times are common. Thus the
typical case is a weak source and
large count numbers. The additional request of a weak source
is expressed by the condition $r \ll 1$.
In this limit 
the two significances actually do agree.

\subsection{Li \& Ma}
Expanding the result in equation~(\ref{limar})
up to the second order with respect to $r$ at $r=0$ gives
\begin{equation}
S_{\mathrm{LM}} \approx 
r\sqrt{\frac{N_{\mathrm{BG}}}{\alpha+1}} \, \left(
1-\frac{2\alpha+1}{6\,(\alpha+1)}\, r 
\right) \,\,.
\label{limaapprox}
\end{equation}
The expansion is done up to the order in which we encounter
a difference to the Bayesian significance. Equation~(\ref{limaapprox}) 
is useful for small values of $r$. The first order term 
is sufficient if one requires
that the second order term is small compared to the leading order.
This gives the condition of how weak the source must be in that case:
\begin{equation}
r \ll \frac{6\,(1+\alpha)}{1+2\,\alpha} \,\,.
\label{ncondlm}
\end{equation}

\subsection{Bayes}
Expanding the Bayesian result for large $N_{\mathrm{BG}}$ - hence $S_2$ 
in equation~(\ref{s2}) - up to the second order with respect to $r$ at $r=0$:
\begin{equation}
S_{\mathrm{B}} = S_2 \approx r\sqrt{\frac{N_{\mathrm{BG}}}{\alpha+1}} \, \left(
1-\frac{1}{4}\, r 
\right) \,\,.
\end{equation}
The first order is sufficient if $r \ll 1/4$. 

\subsection{Comparison}
To first order in $r$, the formula given by Li \& Ma agrees with the
Bayesian result. The difference between the two significances is of second order in $r$:
\begin{equation}
S_{2} - S_{\mathrm{LM}} =  r\sqrt{\frac{N_{\mathrm{BG}}}{\alpha+1}} \,\left(
\frac{1}{12} \frac{\alpha-1}{\alpha+1}\, r 
\right) \,\,.
\label{diff}
\end{equation}
The numerical value of the fraction $(\alpha-1)/(\alpha+1)$ is always in $[-1,\,1]$.
Together with the factor $1/12$ one finds therefore that the relative difference in significance
is typically an order of magnitude smaller than the value of $r$. For $\alpha=1$ 
this relative difference is of order $r^2$. 
This shows that in the case of large count numbers and a weak source the Bayesian result
and the formula given by Li \& Ma are very close to each other. 

Interestingly the correction due to the limited definition region (second term
in equation~(\ref{sbapprox})) is often numerically more important
than the intrinsic difference between the two results as given by formula~(\ref{diff}).
For the case of $r=0.1$, $\alpha=0.3$ and
a typical significance of $3\sigma$ the difference according to equation~(\ref{diff})
is only of order 0.4\%, whereas the limited definition region changes the significance by
6.9\%. The correction by the restricted definition region
is more important than the intrinsic difference 
given by the mathematically different treatment
as long as
\begin{equation}
S^2 \cdot r < 12\,\ln 2 \cdot \left| \frac{\alpha+1}{\alpha-1}\right| \,\,.
\label{rcond}
\end{equation}
This case is relevant since the actual limit of large count
numbers is hard to reach and it quickly leads to
significances which are so high that one could not doubt
the existence of a source.
If condition~(\ref{rcond}) is fulfilled the difference between the two significances
is dominated - technically speaking - by the
definition region. 
The interesting point is that an unrestricted definition
region would allow a source with negative intensity.
Here, physics tells us that a
source can only increase the count number since the source does not interfere
with the background.
In other words: An intensity always has a value $\ge 0$.
One sees how Bayesian statistics
allows us to take into account a-priori
knowledge via the definition region.
In classical statistics a-priori 
knowledge is not taken into account.  
Implicitely the intensity parameter of the source is completey 
free in $]-\infty,\infty[$. 

\begin{figure}
\includegraphics[width=12cm]{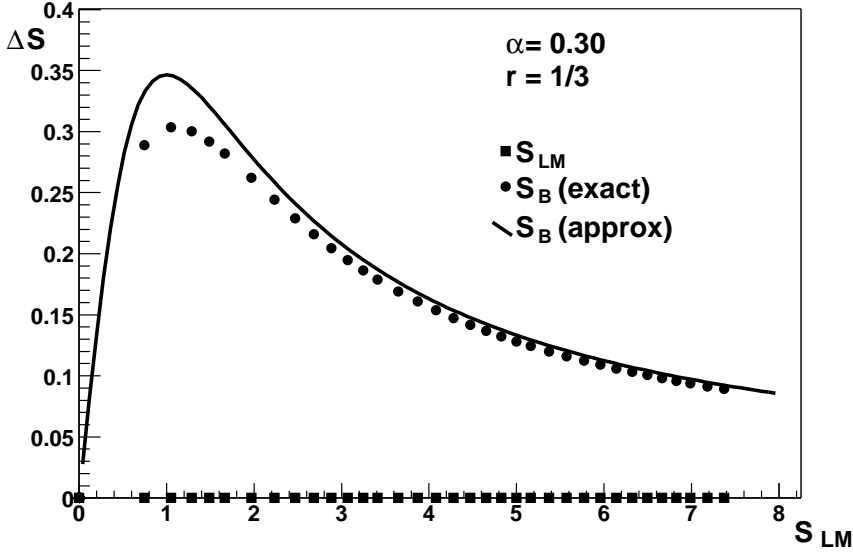}
\caption{~Difference $\Delta S = S - S_{\mathrm{LM}}$ for $\alpha=0.3$, $r=1/3$ 
          as a function of $S_{\mathrm{LM}}$.
	   The significance $S=S_{\mathrm{B}}$ (dots) is moderately
	   higher than the one given by Li \& Ma (squares). The curve
	   shows $S=S_{\mathrm{B}}$ as calculated from the approximation
	   given in equation (\protect{\ref{sbapprox}}). Note that the Bayesian procedure 
	   can only be evaluated for integer count numbers, not allowing for 
	   a continuous coverage on the $S_{\mathrm{LM}}$-axis. The
	   off count number $N_{\mathrm{off}}$ varies from 0 to 2000
	   for $S_{\mathrm{LM}}$ from 0 to 7.5}
\label{fig3}   
\end{figure}

\section{Conclusions}
The decision about a signal in the presence of background has been considered
by Li \& Ma in the framework of classical statistics. We have presented the 
Bayesian treatment of the same problem. This yields a complete solution which
is not restricted to large count numbers. The Bayesian significance
is correct for any $N_{\mathrm{on}}$, $N_{\mathrm{off}}$. \\
We compared the significance by Li \& Ma with the Bayesian
one in the limit of large count numbers. This was dictated by the fact that
Li \& Ma have formulated their expression for that limit. It turns out that
classical statistics and Bayesian inference generally yield different results.
They agree, however, in the limit of large count numbers and a weak source.\\
There are interesting cases where the limit of large count numbers is 
not fully reached. Then an accurate representation of the Bayesian significance
requires a correction of order $N^{-1/2}$ as compared to the leading
term which is of order $N^{1/2}$. There is no room for it in the argument
of Li \& Ma. The correction is due to the fact that a physical intensity parameter
cannot have negative values. Bayesian inference takes care of this piece of
prior knowledge. 

\appendix

\section{Calculation of measures}
\label{calculationofmeasures}
The evaluation of equation (\ref{jeffrey}) is easy using the expectation values
for the respective distribution. For the Poisson distribution one has
\begin{eqnarray}
\left< n \right>_{\mathrm{P}} & = & \lambda \,,\nonumber \\
\left< n^2 \right>_{\mathrm{P}}  & = & \lambda^2 + \lambda \,.
\label{exppoisson}
\end{eqnarray}
For the binomial distribution the expectation values are
\begin{eqnarray}
\left< n \right>_{\mathrm{B}} & = & N \, \lambda \,,\nonumber \\
\left< n^2 \right>_{\mathrm{B}}  & = & N \, \lambda \, ((N-1)\,\lambda +1) \,.
\label{expbinomial}
\end{eqnarray}
The measure of the Poisson distribution is therewith:
\begin{eqnarray}
\ln p_{\mathrm{P}}(n|\lambda)&=&n \, \ln \lambda - \lambda \,,\nonumber \\
\partial_{\lambda} \ln p_{\mathrm{P}}(n|\lambda)&=&
\frac{n}{\lambda} - 1 \,, \nonumber \\
\left< \left( \partial_{\lambda} \ln p_{\mathrm{P}}(n|\lambda) \right)^2\right>_{\mathrm{P}}
&=&1-2+\frac{1+\lambda}{\lambda} = \frac{1}{\lambda} \,,\nonumber \\
\mu_{\mathrm{P}}(\lambda)&=&\lambda^{-1/2} \,.
\label{measurepoisson}
\end{eqnarray}
The measure of the binomial distribution is
\begin{eqnarray}
\ln p_{\mathrm{B}}(n|\lambda;N) &=& \ln {N \choose n} +
n \ln \lambda + (N-n) \ln (1-\lambda) \,,\nonumber \\
\partial_{\lambda} \ln p_{\mathrm{B}}(n|\lambda;N)&=& 
\frac{n}{\lambda}-\frac{N-n}{1-\lambda} \,,\nonumber \\
\left< \left(\partial_{\lambda} \ln p_{\mathrm{B}}(n|\lambda;N) 
\right)^2 \right>_{p_{\mathrm{B}}} &=&
\frac{N}{\lambda (1-\lambda)} \,,\nonumber\\
\mu_{\mathrm{B}}(\lambda)&=&\left(\frac{N}{\lambda (1-\lambda)}\right)^{1/2}\,.
\label{measureofbinomial}
\end{eqnarray}
The conditional measure $\mu_0(\lambda_{\mathrm{off}}|\lambda_{\mathrm{s}})$ needed in
equation (\ref{q})
is calculated in the same way, using the expectation values for
the Poisson distribution:
\begin{samepage}
\begin{eqnarray}
\ln p_0(N_{\mathrm{on}}, N_{\mathrm{off}}| \lambda_{\mathrm{s}}, \lambda_{\mathrm{off}}) &=&
N_{\mathrm{on}} \ln (\lambda_{\mathrm{s}} + \alpha \lambda_{\mathrm{off}}) - \lambda_{\mathrm{s}} - \alpha \lambda_{\mathrm{off}} + 
\nonumber \\ &&
N_{\mathrm{off}} \ln \lambda_{\mathrm{off}} - \lambda_{\mathrm{off}} - \ln (N_{\mathrm{off}}!\, N_{\mathrm{on}}!) \,,\nonumber \\
\partial_{\lambda_{\mathrm{off}}} \ln p_0(N_{\mathrm{on}}, N_{\mathrm{off}}| \lambda_{\mathrm{s}}, \lambda_{\mathrm{off}})
& = & \frac{\alpha \, N_{\mathrm{on}}}{\lambda_{\mathrm{on}}} - \alpha
+ \frac{N_{\mathrm{off}}}{\lambda_{\mathrm{off}}} - 1 \,,\nonumber \\
\left< \left( \partial_{\lambda_{\mathrm{off}}} \ln p_0(N_{\mathrm{on}}, N_{\mathrm{off}}| \lambda_{\mathrm{s}}, \lambda_{\mathrm{off}})
\right)^2 \right>_{p_0} &=& \frac{\alpha^2}{\lambda_{\mathrm{on}}} + \frac{1}{\lambda_{\mathrm{off}}}
\,,\nonumber \\
\mu_0(\lambda_{\mathrm{off}}|\lambda_{\mathrm{s}}) &=& \left( 
\frac{\alpha^2}{\alpha \lambda_{\mathrm{off}} + \lambda_{\mathrm{s}}} + 
\frac{1}{\lambda_{\mathrm{off}}} \right)^{1/2} \,.
\label{measuremu0}
\end{eqnarray}
 \end{samepage}
\section{Check if the minor model is proper}
\label{evalq}
It has to be checked whether the model $q_0$ in equation (\ref{q}) is proper. 
Thus one has to evaluate
\begin{eqnarray}
\sum_{N_{\mathrm{on}},\atop N_{\mathrm{off}}} q_0(N_{\mathrm{on}}, N_{\mathrm{off}}| \lambda_{\mathrm{s}}) & = & 
\sum_{N_{\mathrm{on}},\atop N_{\mathrm{off}}} \int_0^{\infty} p_0(N_{\mathrm{on}}, N_{\mathrm{off}}| \lambda_{\mathrm{s}}, \lambda_{\mathrm{off}}) \, 
\mu_0(\lambda_{\mathrm{off}}|\lambda_{\mathrm{s}}) \,d\lambda_{\mathrm{off}} \nonumber\\
& = & \int_0^{\infty} \mu_0(\lambda_{\mathrm{off}}|\lambda_{\mathrm{s}}) \,d\lambda_{\mathrm{off}} \nonumber\\
&=& \int_0^{\infty} \left( 
\frac{\alpha^2}{\alpha \lambda_{\mathrm{off}} + \lambda_{\mathrm{s}}} + 
\frac{1}{\lambda_{\mathrm{off}}} \right)^{1/2} \,d\lambda_{\mathrm{off}} \,.
\end{eqnarray}
This integral diverges and hence $q_0(N_{\mathrm{on}}, N_{\mathrm{off}}| \lambda_{\mathrm{s}})$
is an improper model.

\section{Transformation to a proper model}
The transformation from the original parameters $(\lambda_{\mathrm{s}},\lambda_{\mathrm{off}})$ 
to the new ones $(\omega,\Lambda)$ is calculated in a few lines:
\begin{samepage}
\begin{eqnarray}
p_0(N_{\mathrm{on}}, N_{\mathrm{off}}| \lambda_{\mathrm{s}}, \lambda_{\mathrm{off}}) & \stackrel{(\ref{p0})}{=} 
& \frac{\lambda_{\mathrm{on}}^{N_{\mathrm{on}}}}{N_{\mathrm{on}}!} \, e^{-\lambda_{\mathrm{on}}} \, 
\frac{\lambda_{\mathrm{off}}^{N_{\mathrm{off}}}}{N_{\mathrm{off}}!} \, e^{-\lambda_{\mathrm{off}}} \nonumber\\
& = & \frac{(\Lambda \, \omega)^{N_{\mathrm{on}}}}{N_{\mathrm{on}}!} \, 
\frac{(\Lambda (1 - \omega))^{N_{\mathrm{off}}}}{N_{\mathrm{off}}!} \, e^{- \Lambda} \nonumber \\
& = & \Lambda^{N} \, e^{- \Lambda} \, \frac{1}{N_{\mathrm{on}}! \, N_{\mathrm{off}}!} \, \omega^{N_{\mathrm{on}}} (1-\omega)^{N_{\mathrm{off}}} \nonumber\\
& = & \frac{\Lambda^{N} \, e^{- \Lambda}}{N!} \, \nonumber
{N \choose N_{\mathrm{on}}} \, \omega^{N_{\mathrm{on}}} (1-\omega)^{N - N_{\mathrm{on}}} \\
& \stackrel{(\ref{poisson}),(\ref{binomial})}{=} & 
p_{\mathrm{P}}(N|\Lambda) \cdot p_{\mathrm{B}}(N_{\mathrm{on}}|\omega;N) \,.
\label{p0t}
\end{eqnarray}
\end{samepage}
\section{Uniqueness of the solution}
\label{proof}
The first derivative of the model
$p_{\mathrm{B}}$ from equation (\ref{binomial}) is
\begin{equation}
p'_{\mathrm{B}}(\omega | N_{\mathrm{on}}; N) = 
\left( \frac{N_{\mathrm{on}}}{\omega} - \frac{N_{\mathrm{off}}}{1-\omega}  \right)
\, p_{\mathrm{B}}(\omega | N_{\mathrm{on}}; N) \,.
\label{p1der}
\end{equation}
It vanishes at
\begin{equation}
\omega_0 = \frac{N_{\mathrm{on}}}{N} \,.
\end{equation}
The value of the second derivative at $\omega_0$ is
\begin{equation}
p''_{\mathrm{B}}(\omega_0 | N_{\mathrm{on}}; N) = - \frac{N^3}{N_{\mathrm{off}}\,N_{\mathrm{on}}} < 0 \,.
\end{equation}
Hence, $p_{\mathrm{B}}$ has a single maximum and no minima. 
Therefore for each $\omega_1 \ne \omega_0$
one has exactly one other $\omega_2$ for which equation~(\ref{eqnbin}) holds.
Thus one has a unique solution $\omega_{\mathrm{up}} \ne \omega_{\mathrm{min}}$
in equation~(\ref{eqnbin}). 

\section{Approximation to the posterior distribution}
\label{gaussapprox}
The result of the transformation of $p_{\mathrm{B}}(\omega)$ to 
the parameter $\phi=\arcsin(\sqrt{\omega})$ is
\begin{equation}
p_{\phi}(\phi) \sim (\sin \phi)^{2N_{\mathrm{on}}} \, (\cos \phi)^{2N_{\mathrm{off}}} \,.
\end{equation}
The approximation is achieved by expanding the logarithm of the distribution
around its maximum and taking the exponential of the result.
Using $\gamma = N_{\mathrm{on}}/N_{\mathrm{off}} = \alpha\,(1+r)\,$ the maximum is at
\begin{equation}
\phi_0 = \arctan \left( \sqrt{\gamma}\,\right) \,.
\end{equation}
The expansion up to second order is
\begin{equation}
\ln p_{\phi}(\phi) = C_1(N_{\mathrm{on}},N) - 2 N \, \left( \phi - \phi_0 \right)^2 
+O[\phi^3] \,.
\label{lnp}
\end{equation}
Hence, one has
\begin{equation}
\tilde{P_2} = \frac{1}{C_2}
\exp\left(-2 N (\phi-\phi_0)^2 \right)
\end{equation}
With the normalization constant
\begin{equation}
C_2= \sqrt{\frac{\pi}{2N}}
\end{equation}
the distribution $P_2$ is normalized in $]-\infty,\infty[$.

\section{Mathematica script to evaluate Bayesian significance}
\label{script}
Although we cannot give a close formula for the Bayesian significance,
we can show a short Mathematica script which calculates the significance
as given in equation~(\ref{result}).
\begin{samepage}
\begin{verbatim}
data = {   a -> 0.25, 
         non -> 16, 
        noff -> 10 };

n = non + noff; 
b = non/noff;
wmin = a/(1 + a);
pBin[x_, n_, non_] := Binomial[n, non]x^non(1 - x)^(n - non); 
pRaw[x_, n_, non_] := pBin[x, n, non] (Sqrt[n/x(1 - x)]); 
norm = Integrate[pRaw[x, n, non], {x, wmin, 1}];
p[x_] := pRaw[x, n, non]/norm; 
rule =  FindRoot[Evaluate[(1 - w) (1 + a)  == (wmin/w)^b /.data], 
                 {w, wmin/a, non/n, 1}/.data];
i[w0_, w1_] := Integrate[p[w], {w, w0, w1}, GenerateConditions -> False]; 
temp = Evaluate[(i[wmin, w /. rule]) /. data]; 

Print["Sigma (Bayes): "];
sigma = InverseErf[temp] Sqrt[2]
\end{verbatim}
\end{samepage}

\clearpage

\end{document}